\documentclass[twocolumn,showpacs,preprintnumbers,pre,fleqn]{revtex4}
\usepackage{epsfig}
\usepackage{graphicx}
\usepackage{amsfonts}
\newcommand{\ig}{\includegraphics}
\newcommand{\ct}{\cite}

\newcommand{\beas}{\begin{eqnarray*}}
\newcommand{\eeas}{\end{eqnarray*}}
\newcommand{\be}{\begin{equation}}
\newcommand{\ee}{\end{equation}}
\newcommand{\ba}{\begin{eqnarray}}
\newcommand{\ea}{\end{eqnarray}}

\begin{document}
                                                                                

\title{A random fiber bundle with many discontinuities in the threshold distribution}
\author{Uma Divakaran}
\email{udiva@iitk.ac.in}
\author{Amit Dutta}
\email{dutta@iitk.ac.in}
\affiliation{Department of Physics, Indian Institute of Technology Kanpur - 208016, India}
\date{\today}
\begin{abstract}

We study the breakdown of a random fiber bundle model (RFBM) with $n$-discontinuities in the
threshold distribution using the global load sharing
scheme. In
other words, $n+1$ different classes of fibers identified on the basis of their threshold
strengths are mixed such that the strengths of the fibers in the $i-th$ class
are uniformly distributed between the values $\sigma_{2i-2}$ and $\sigma_{2i-1}$ where $1 \leq i \leq n+1$. Moreover, there is a gap in the threshold distribution between $i-th$ and $i+1-th$ class.
We show that although the critical stress depends on the parameter values of 
the system, the critical exponents are identical to that obtained in the
recursive dynamics of a RFBM with a  uniform
distribution and global load sharing.  The avalanche size
distribution (ASD), on the other hand,  
shows a non-universal, non-power law behavior
for smaller values of avalanche sizes which becomes prominent only when a 
critical distribution is approached.
We establish that the behavior of the avalanche size distribution for an 
arbitrary $n$ is qualitatively similar to a RFBM with a single discontinuity  
in the threshold distribution ($n=1$), especially when the density and the 
range of threshold values of fibers belonging to strongest ($n+1$)-th class 
is kept identical in all the cases.

-

\end{abstract}
\pacs{46.50. +a, 62.20.Mk, 64.60.Ht, 81.05.Ni}  
\maketitle
\section{Introduction}
What causes fracture of materials in nature? Are there any precursor that signals
the imminence of a
complete breakdown so that we can avoid them taking place? 
These are few of the many questions that physicists 
and engineers are looking at now a days  to explore the fracture 
dynamics of heterogeneous materials\ct{benguigui,zapperi97}. A prior knowledge of 
failure properties of such materials are of extreme importance for problems related to physics
of breakdown,
material science  as well as architectural, mechanical
and textile engineering.  The simplest of all the attempts 
is a model of fibers with randomly distributed threshold strengths known as
the random fiber bundle model. 

In a random fiber bundle model (RFBM) \ct{silveria,hemmer,aval,moreno93,andersen,moreno00, reviewchak,dynamic,prl,kim05,kun06,divakaran071,divakaran07,pradhan07,hidalgo08}, 
fibers with stochastically distributed values of threshold strengths (i.e., the
maximum external stress a fiber can withstand) are clamped at both the ends. Since the 
threshold of a fiber
depends crucially upon the presence of defects in that particular fiber, it is indeed useful to assign random
threshold strength to each fiber of the bundle. The threshold strengths however are chosen from a  given distribution, usually approximated by a Weibull or uniform 
distribution. Under the application of a weak external load, the fibers with threshold  lying
below the applied stress, break and the resulting additional load is distributed among the intact fibers using
a load sharing rule. This redistribution of stress causes further failures and  the dynamics stops when the system 
 reaches a fixed point at which no further failure takes place. To resume the recursive dynamics, the external load is further increased 
 to break the next weakest intact fiber.  The process continues until the 
complete breakdown of the entire bundle.  

In this work, we shall use the global load sharing (GLS) rule where the
additional load generated due to the breaking of a fiber is shared equally by all the intact fibers of the bundle.
In the GLS scheme, the breakdown of the fiber bundle can be interpreted as a continuous phase transition with
well defined critical stress and critical exponents \ct{andersen,moreno00,reviewchak,dynamic}. For a RFBM with GLS, 
it has also been established that 
  the distribution $D(\Delta)$ of an avalanche  of size $\Delta$, defined as the number of fibers broken between two successive loadings, satisfies a universal power-law $D(\Delta) \sim \Delta^{-5/2}$ in the limit of
$\Delta \to \infty$ \ct{hemmer,aval}. In a recent work, Pradhan, Hansen and Hemmer \ct{prl} showed that in the vicinity of the critical
distribution where the average external load on the bundle is maximum, the avalanche size distribution
shows a crossover to a new power-law behavior given by $D(\Delta) \sim \Delta^{-3/2}$  for small values of $\Delta <<\Delta_c$ while
for $\Delta >> \Delta_c$, the $\Delta^{-5/2}$ behavior is recovered. The characteristic size $\Delta_c$ around which the
crossover occurs diverges in a power-law fashion as the critical distribution is approached and at the critical distribution
the $\Delta^{-3/2}$ behavior is observed for the entire range of $\Delta$.
 
\begin{figure}[h]
\includegraphics[height=1.9in]{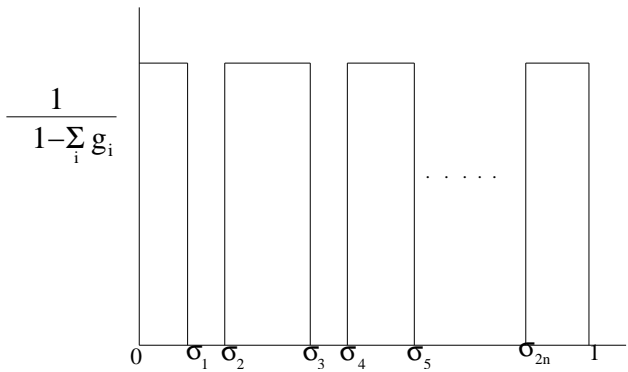}
\caption{Mixed Uniform Distribution with  $(n+1)$ classes of fibers  and $n$-gaps
or discontinuities. }
\end{figure}

In a recent  paper, the authors \ct{divakaran07} investigated the robustness of the above universal power-law behavior of
the avalanche size distribution by introducing a discontinuity in the threshold 
distribution $\rho(\sigma_{th})$ which is given by 
\begin{eqnarray}
\rho(\sigma_{th}) &=& \frac {1}{1-(\sigma_{2}-\sigma_{1})}~~~~~~
0<\sigma_{th}\leq \sigma_{1}\nonumber\\
&=&0~~~~~~~~~~~~~~~~~~~~~~\sigma_{1}<\sigma_{th}<\sigma_{2}\nonumber\\
~~~~~~~~~~&=&\frac {1}{1-(\sigma_{2}-\sigma_{1})}~~~~~\sigma_{2}\leq\sigma_{th}\leq 1.
\end{eqnarray}
Hence, two types of fibers (weaker and stronger)
separated by a gap in their threshold distributions coexist in the 
same bundle with more than half of the fibers belonging to
the stronger class. It has been established that there exists a 
non-universal non-power law behavior in the avalanche
size distribution for small $\Delta$ which crosses over to the universal 
behavior $\Delta^{-5/2}$ in the asymptotic 
limit of $\Delta$. Most interestingly, it was pointed out that the non-universality becomes prominent only when a critical
distribution is approached. The threshold distribution given in Eq.~1 
with $\sigma_2=0.5$ is the
critical distribution of a RFBM with single discontinuity\ct{divakaran07},
because as soon as the redistributed stress reaches 0.5, the bundle is 
critical and breaks down completely with an infinitesimal increase 
in the external load.
At a critical distribution, however, there is a crossover from the non-universal behavior
to a power-law behavior with $ \Delta^{-3/2}$ for  large $\Delta$. A recent study of an
infinite gap generalization of the above model predicts
a new exponent (=$9/4$) of the avalanche size distribution\ct{hidalgo08}.

 The natural question that remains is what would be  the effect of many
 such discontinuities
on the avalanche size distribution? We address this issue in the present
communication where
we study a RFBM with $n$-discontinuities in the threshold
distribution and investigate its effect on the  critical behavior and the 
avalanche size distribution.
The paper is organized as follows: In Section II, we introduce
the model and derive 
the critical stress and exponents using the recursive dynamics approach . The results on the avalanche size distribution and comparison with the $n=1$ case studied previously are presented in section III.  We make concluding
comments  in
section IV.

\section{The model and the recursive dynamics}

The threshold distribution of a RFBM with  $n$-discontinuities studied in this paper is  shown in Fig.~1. The threshold range of the weakest and
the strongest class of fibers is from $0$ to $\sigma_1$ and 
$\sigma_{2n}$ to $1$, respectively.
The mathematical form of the normalized threshold distribution is given by 
\begin{eqnarray}
\rho(\sigma_{th}) &=& \frac 1 { 1- \sum_{i=1}^n (\sigma_{2i}-\sigma_{2i-1})}\nonumber\\
&=& ~~~~~~
\frac {1}{1- \sum_{i=1}^n {g_i} }~~~{\rm for}~~
\sigma_{2i-2}<\sigma_{th}\leq \sigma_{2i-1}\\
&=&0~~~~~~~~~~~~~~~~~~~~~~{\rm otherwise} \nonumber\\
\end{eqnarray}
where $1 \leq i \leq n+1$, $g_i =\sigma_{2i}-\sigma_{2i-1}$ and $ \sigma_0 =0$, $\sigma_{2n+1} =1$. 
The above distribution suggests that $(n+1)$ different classes of fibers with
the threshold of the fibers of the $i$-th class ranging from $\sigma_{2i-2}$ to $\sigma_{2i-1}$ are mixed.
At the same time the restriction  $0 < \sigma_1 < \sigma_2...<\sigma_{2i-1}
<\sigma_{2i} <1$ ensures the existence of a gap or discontinuity given
by $g_i$ in the threshold strengths between $i$-th and $(i+1)$-th class of fibers.  We also assume that the threshold values of the  $i$-th class of fibers are uniformly distributed 
  within the range $\sigma_{2i-2}$ to $\sigma_{2i-1}$  for all $i$, a condition that
leads to 
a set of  $(n+1)$ additional  relations connecting the parameters of the system in the following
way. Distributing a fraction $f_i$ of total number of fibers
to the $i$-th class (with $\sum_i^{n+1}f_i=1$) and using the uniformity condition mentioned above 
we get

\ba
f_{i} = \frac {\sigma_{2i-1}-\sigma_{2i-2}} {1- \sum_{i=1}^n {g_i} }
\ea
such that $\sigma_0=0$, $\sigma_{2n+1}=1$ and once again $1\leq i \leq n+1$
The model therefore involves $(2n)$ values of $\sigma_i$'s, $(n+1)$-values of 
density $f_i$, the 
conditions given by Eq.~4 along with the additional restrictions  
$\sum_i^{n+1} f_i=1$. Therefore, the total number of free variables that can be 
chosen independently reduces to $2n$. It should also be noted
that since ${1- \sum_{i=1}^n g_i }$ is always less than unity, we must have  
$(\sigma_{2i-1}-\sigma_{2i-2}) < f_i$ for all $i$.

We shall estimate the critical stress of the above model within the framework of a recursive dynamics
and global load sharing. The fibers belonging to all $(n+1)$-classes cooperatively participate in sharing the
additional load arising due to the breaking of the weaker fibers. 
The 
fraction of unbroken 
fibers after a time step $t+1$, denoted by $U_{t+1}$,  is related to $U_t$  through the relation
\cite {dynamic, divakaran07}
\begin{equation}
U_{t+1} = 1-P(\sigma_{t})=1-P(\frac{\sigma}{U_{t}})
\end{equation}
\noindent where $P(\sigma_t)$ is the fraction of broken fibers with the applied stress $\sigma$ and
redistributed stress $\sigma_t$, and is given as
$$P(\sigma_t)=\int_0^{\sigma_t}\rho(\sigma_{th})d\sigma_{th}.$$
Similarly, the redistributed stress after a time step $(t+1)$ satisfy the recursive relation
\be
\sigma_{t+1} =\frac{\sigma}{U_{t+1}} =
\frac {\sigma} {(1-P(\sigma_t))}.
\ee
 The fixed point solution for $U$ ($=U^*$)  at which no further failure takes place
can be obtained by solving the above recursive
relations (5) and (6)\ct{dynamic}. Assuming that the redistributed stress at some instant $t$, $\sigma_t$ exceeds  $\sigma_{2n}$ (i.e, when the redistributed
stress initiates the breaking of the $(n+1)$-th class fibers), we obtain
\ba
U_{t+1} &=& 1-P(\frac{\sigma}{U_{t}}) \nonumber \\
 &=& 1-[\sum_{i=1}^n\frac{\sigma_{2i-1} -\sigma_{2i-2}}{1- \sum_i^n g_i} + \frac{1}{1- \sum_i^n g_i}(\frac{\sigma}{U_{t}}-\sigma_{2n})]\nonumber
\end{eqnarray}
so that we get the fixed point solution
\be
U^{*} = \frac {1}{2(1-\sum_i^n g_i)} [1 + \sqrt
{1-\frac{\sigma}{\sigma_{c}}]}
\ee
Along the same line of arguments, we find the redistributed stress at the fixed 
point 
\be
\sigma^{*} = \frac 1 {2} - \frac 1 {2} \sqrt { 1 - \frac {\sigma}{\sigma_c}}.
\ee 
Using Esq.~ (7) and (8) we find that the critical stress is given by
 \be 
\sigma_{c} = \frac {1} {4[1-(\sum_i^n g_i)]}. 
\ee
which is the applied load per fiber at which half of the fibers break.
When the external load exceeds $\sigma_c$ even by an infinitesimal amount, there is no real
solution 
of $U^{*}$ and $\sigma^{*}$ which signals the complete break down of the bundle.

The critical stress of the mixed model  varies with the gaps
$g_i$  and  in the limit $g_i \to 0$
for each $i$, we retrieve the critical stress  $\sigma_c =1/4$ for a RFBM 
with uniform distribution \ct{dynamic}. We also get back the result of single discontinuity
case i.e., $n=1$ \ct{divakaran07} if only  $g_1\neq 0$. The equation (8) also shows
that the redistributed stress attains the
maximum value of 0.5 at the critical  external stress $\sigma_c$. 

Calling the redistributed stress as x from now onwards,
we consider the constitutive equation, 
 $F(x) = N x (1 - P(x)$ where $ F(x)$  is the average external 
load when the redistributed stress is $x$.  The load $F(x)$ maximizes 
when the redistributed stress $x$ is equal to $0.5$. Since the maximum 
value of the redistributed stress is equal to 0.5, 
$\sigma_{2n}$ must be less
than 0.5  so that some fibers from
$(n+1)$-th class also fail at the critical point. The constraint equations (4) suggest 
that for recursive dynamics to hold good,
 more than half of the fibers must belong to the $(n+1)$-th class with 
thresholds lying between
$\sigma_{2n}$ to $1$. We therefore define the threshold distribution 
$\rho(\sigma_{th})$
(Eq.~2), along with the constraint conditions and  
$\sigma_{2n} = 0.5$ as the critical distribution.
It can also be shown in a straightforward way that the order parameter 
exponent ($\beta$) and the susceptibility
exponents ($\gamma$) stick to the mean field (GLS) values with $\beta =\gamma=1/2$, 
even in the presence of an arbitrary number of 
discontinuities provided uniformity condition is satisfied
and hence the critical behavior remains unaltered although there is an appreciable change in the 
critical stress.

\section{Avalanche Size Distribution (ASD)}
We shall now turn our attention towards the ASD of RFBM in the presence
of many discontinuities in the threshold distribution which is the key 
point of our study.
We show below that the discontinuities have a non-trivial affect on the 
ASD when a critical distribution is approached. 
We also argue that a situation
 with many discontinuities is qualitatively similar to the single-discontinuity case. The scenario is established considering a special case with $n=2$, 
where $g_1 = \sigma_2 -\sigma_1$ and $g_2=\sigma_4 -\sigma_3$ so that 
three classes of fibers with range
of thresholds lying between ($0$ to $\sigma_1$), ($\sigma_2$ to $\sigma_3$) and
($\sigma_4$ to $1$), respectively, coexist in the bundle while $f_1$, $f_2$ and
$f_3$ are the corresponding densities.

Below is shown some of the allowed distributions which 
satisfy the restrictions mentioned in Eq.~4 and case 4 refers to a critical distribution.

\begin{center}
\begin{tabular}{|c|c|c|c|c|c|c|c|}\hline
Case &~~~$f_1$~~~&~~~$f_2$~~~&~~~ $\sigma_{1}$~~~~&~~~~$\sigma_{2}$ ~~~~ & ~~~~$\sigma_{3}$~~~&~~~$\sigma_4$~~~&~~~$\sigma_c$\\\hline
1 &0.10  &.20  &0.08  &0.16 &0.32 &0.44 &0.31\\\hline
2 &0.15  &0.25  &0.135  &0.15 &0.375 &0.46&0.27\\\hline
3 &0.05  &0.05  &0.04  &0.16 &0.20 &0.28 & 0.31\\\hline
4 &0.10  &0.20  &0.07  &0.16 &0.30 &0.50 &0.35\\\hline
5 &0.10&0.2&0.08&0.26&0.42&0.44&0.31\\\hline
6 &0.20&0.15&0.16&0.2&0.32&0.48&0.31\\\hline

\end{tabular}
\end{center}

\begin{figure}[h]
\ig[height=3.6in,width=3.6in]{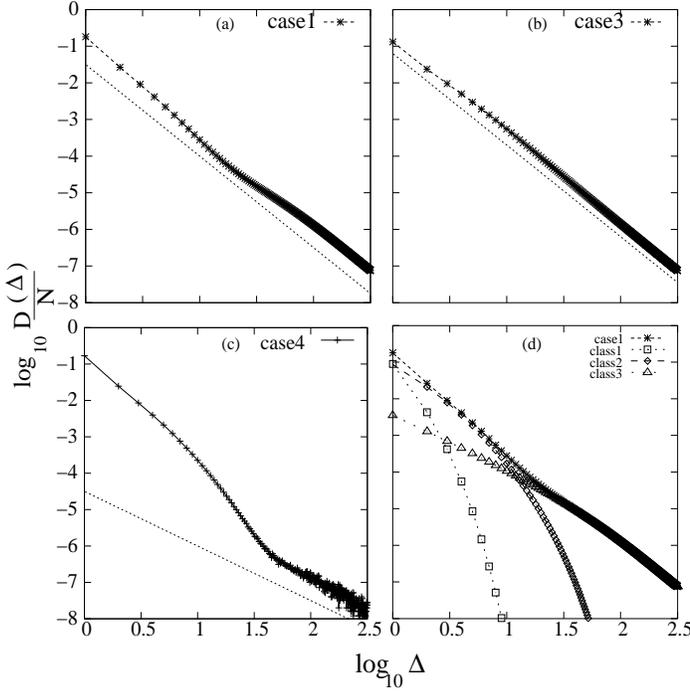}
\caption{
Fig.~(2a) corresponds to the ASD for case1 of the table 
where a crossover from nonuniversal to the universal $5/2$ behavior
 is observed since
$\sigma_4$ is close to $0.5$. 
(2b) shows case3 of the table where 
no non-universal behavior is observed except for small values of $\Delta$. 
ASD for a critical 
distribution (case4) is shown in Fig.~2(c) where there is a crossover to the "3/2" behavior
for large $\Delta$. 
The dotted line in Fig~2a and 2b has a slope -5/2 whereas that in Fig~2c has
a slope -3/2.
In 2(d), we show $D_1(\Delta)$, 
$D_2(\Delta)$, $D_3(\Delta)$ and the total ASD for the case1 
to compare their relative magnitudes.}
\end{figure}

We shall now  generalize the results of  Hemmer and Hansen \cite{hemmer} to study
the avalanche behavior of the 
 $n=2$-discontinuity model. The general expression for the avalanche size distribution
with GLS is given as
\ba
\frac{D(\Delta)}{N}&=&\frac{\Delta^{\Delta-1}}{\Delta!}\int_0^{x_c}dx\rho(x)
 \frac {(1-a(x))}{a(x)} \nonumber\\ +
&~&~~~~~~~~~~~~~~~\times\exp(\{-a(x)+\ln a(x)\} \Delta)
\ea
where $x$ is the redistributed stress and the upper limit of the integration 
($x_c$)
is the redistributed stress at the critical point. Also,  
$a(x) = x\rho(x)/1-P(x)$ is the number of fibers that break as a result of 
breaking a fiber with threshold
strength $x$ by applying an external load $x/\{(1-P(x))N\}$.

\noindent Let us first consider the breaking of  fibers belonging to class 1 
with threshold values
uniformly lying between $0$ to $\sigma_1$, this contribution ${D_1(\Delta)}$ is given by

\ba
{D_1(\Delta)}&=&\frac{\Delta^{\Delta-1}}{\Delta!}\int_0^{\sigma_1} dx \frac {1}{1-g_1 -g_2} \frac {(1-a(x))}{a(x)} \nonumber\\
&~&~~~~~~~~~~~~~~~\times\exp(\{-a(x)+\ln a(x)\} \Delta)
\ea
where $a(x) = x/(1-g_1-g_2 -x)$.  The maximum contribution of this integral 
is at $a(x)=1$ or $x = (1-g_1 -g_2)/2 $ which exceeds $\sigma_1$, i.e., lies
 beyond the range of integration and we can not employ the method of the 
saddle point integration. However, $a(x)$ is a monotonically increasing 
function of $x$ up to $\sigma_1$ and we therefore get the  
maximum contribution  when
$x$ reaches the upper limit of the integration $\sigma_1$. 
The leading behavior of ${D_1(\Delta)}$  
(with $\Delta! = \exp(-\Delta) \Delta^{\Delta} \sqrt {2\pi \Delta}$), 
is given by
\be
{D_1(\Delta)} \sim \frac {1}{\sqrt {2\pi} (1-g_1 -g_2)} \Delta^{-5/2} x_m^\Delta \exp ((1-x_m) \Delta),
\ee
where $x_m = \sigma_1 /(1-g_1 -g_2 -\sigma_1) = f_1 /(1-f_1)$. Therefore, $D_1(\Delta)$ exhibits a
non-universal decay with increasing $\Delta$ which is more rapid if $f_1 \to 0$. 

Similarly, the leading contribution of the fibers belonging to the class 2 with threshold values ranging from
$\sigma_2$ to $\sigma_3$  and $a(x) = x/(1-g_2-x)$ is found to be
\be
{D_2(\Delta)} \sim \frac {1}{\sqrt {2\pi} (1-g_1 -g_2)} 
\Delta^{-5/2} y_m^\Delta \exp ((1-y_m )\Delta),
\ee
where $y_m = \sigma_3/(1-\sigma_4)$. Therefore, $D_2(\Delta)$ shows a similar non-universal behavior which survives even for relatively higher values of $\Delta$
if $\sigma_3$ approaches $\sigma_4$ and also $\sigma_4 \to 0.5$. The significance of the
above findings is explained below.

Let us now focus on the contribution from the fibers belonging to
class $3$. Using Eq.~(10) with $a(x)=x/(1-x)$ and  
$f_3 = (1-\sigma_4)/(1-g_1 -g_2)$, we get
\ba
{D_3(\Delta)}&=&\frac{1}{\sqrt {2\pi \Delta}} \frac {f_3}{1-\sigma_4}\int_{\sigma_4}^{0.5} dx  \frac {(1-2x)}{x}\nonumber\\
&~& ~~~~~~~~
\times\exp(\{-\frac {x}{1-x} +\ln(\frac {x}{1-x}) \} \Delta)
\ea
Right hand side of Eq.~(14) can be integrated
to obtain,
\be
{D_3(\Delta)}\sim \frac{f_3}{2\sqrt {2\pi} \Delta^{5/2}(1-\sigma_4)} \left (1 - e^{-\frac {\Delta}{\Delta_c}} \right)
\ee
where $\Delta_c \sim (1/2 -\sigma_4)^{-2}$, which diverges in the limit 
$\sigma_4 \to 0.5$. 
Our observations are depicted in Fig.~2, where $D(\Delta)$, obtained by numerical integration and also by simulation
using the weakest fiber approach \ct{hemmer}, is plotted for different threshold distributions. 
It is to be noted that the contribution ${D_3(\Delta)}$ depends only on 
$\sigma_4$ and $f_3$, an observation that leads to an interesting conclusion 
that the contribution to the
total avalanche size distribution coming  from the strongest class of fibers is 
identical for any number of discontinuities
if the fraction of fibers as well as the range of the threshold strength of the 
final block is kept fixed (see Fig.~3). Equation (15) provides two 
limiting power-law behaviors given as

\ba
{D_3(\Delta)}&\sim& \Delta^{-3/2} ~{\rm for} ~\Delta < \Delta_c\nonumber \\
&\sim& \Delta^{-5/2}~{\rm for} ~\Delta > \Delta_c
\ea
Comparing Eqs. (12), (13) and (16), we observe that in the limit of small
$\Delta$, the non-power law contributions from $D_1(\Delta)$ and $D_2(\Delta)$
dominates over the universal "$5/2$" behavior only in the limit of
$\sigma_4 \to 0.5$ when ${D_3(\Delta)} \sim \Delta^{-3/2}$. Otherwise,
${D_3(\Delta)}$ dominates over the non-universal contributions so that
one observes a universal behavior nearly for  the entire range of $\Delta$
though the discontinuities in the distribution always exist. At a critical distribution
however, there is a crossover to the universal behavior $D(\Delta)\sim\Delta^{-3/2}$ for very large $\Delta$ following a large region of non-universality (see Fig.~2c).
This general behavior
is valid for any number of discontinuities including $n=1$ \ct{divakaran07}. 
\begin{figure}[h]
\ig[height=2.2in,width=2.2in]{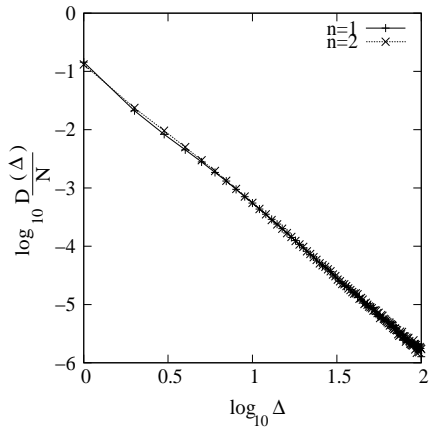}
\caption{Comparison of total avalanche size distribution for same value of $f_{n+1}$
and $\sigma_{2n}$ when $n=1$ and $n=2$ where $n$ is the number of discontinuities.
 Clearly, the two cases overlap in the 
large $\Delta$ region where the final block dominates. Here, $f_{n+1}$=0.9 
and $\sigma_{2n}$=0.28}.
\end{figure}

Let us now concentrate on some interesting limiting situations to investigate
the role of two discontinuities: (i) If the fraction $f_1$ of fibers in class 1 
is small, the contribution $D_1(\Delta) \sim [f_1/(1-f_1)]^{\Delta}$ dies
off rapidly. However, if $\sigma_3$ is large and $\sigma_4$ approaches
0.5, there is a wide region of non-universality in $(D(\Delta)-\Delta)$ behavior
which is solely due to $D_2(\Delta)$ which scales as 
$[\sigma_3/(1-\sigma_4)]^{\Delta}$. (ii) In the other limit, when $f_2 < f_1$ (but $f_2+f_1 <0.5$, as required), $D_1(\Delta)$ do dominate in the small $\Delta$ limit, but for large
$\Delta$ once again it is the contribution of $D_2(\Delta)$, rather the 
larger value of $\sigma_3$ that leads to a prominent non-universal behavior,
$e.g.$, case 6 in table 1. We therefore conclude that the contribution from the weakest
class of fibers is not significant in the large $\Delta$ limit and it is the
higher value of $y_m = \sigma_3/(1-\sigma_4)$ that causes the non-universality to survive
up to relatively higher values of $\Delta$ (Fig. 4)

\begin{figure}[h]
\ig[height=1.9in,width=3.0in]{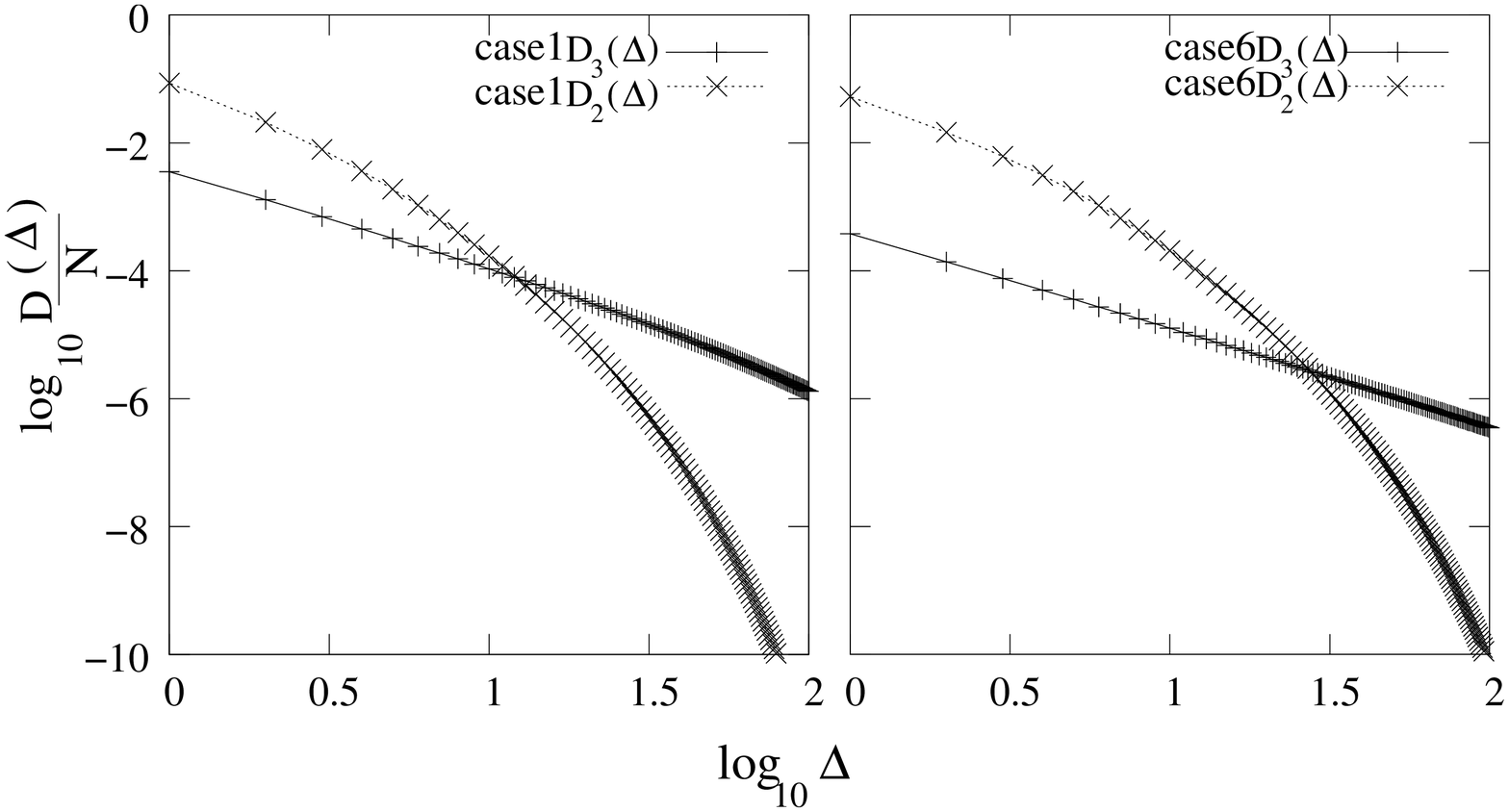}
\caption{The figure shows the comparison of $D_2(\Delta)$ with $D_3(\Delta)$ for case 1
(left panel) and case 6 (right panel). The figure shows that the crossover to universal behavior
occurs at higher value of $\Delta$ if $y_m=\sigma_3/(1-\sigma_4)$ increases.}
\end{figure}

\section{Conclusions}
In conclusion, we have studied a mixed fiber bundle with many discontinuities
in the threshold distribution and GLS  where threshold values of the fibers 
belonging to a particular class are uniformly distributed within the 
specified range. Our 
studies lead to the following conclusions for an arbitrary number of 
discontinuities: (i) The recursive dynamics studies point
to the existence of a critical distribution as defined in the text. 
(ii) There exists a non-universal,
non- power law behavior in the avalanche size distribution for small 
$\Delta$ which
becomes prominent only when a critical distribution is approached, otherwise it is masked by the 
universal behavior except for very small $\Delta$. For asymptotically large
$\Delta$, however, there is always a crossover to the universal 
behavior with $\xi =5/2$
(or $\xi=3/2$ at the critical distribution). The crossover occurs around 
$\Delta=\Delta_c$ where
the contribution from the strongest class of fibers $D_{n+1}(\Delta)$ 
switches from $\xi=3/2$ to
$\xi=5/2$ behavior. 
(iii) $D_{n+1}(\Delta)$ is found to depend only on $f_{n+1}$ and $\sigma_{2n}$
so the contribution of the fibers belonging the strongest class 
(i.e., the behavior of
total $D(\Delta)$ in the limit of large $\Delta$) remains identical for any 
number of discontinuities if $f_{n+1}$ and $\sigma_{2n}$ are kept fixed. 
(iv) We also show that if $\sigma_{2n-1}$ increases and at the same time 
$\sigma_{2n} \to 0.5$, the non-universality survives up to higher
values of $\Delta$.


\begin{thebibliography}{05}

\bibitem{benguigui}
B. K. Chakrabarti and L. G. Benguigui, {\it Statistical Physics of fracture and
Breakdown in Disordered Systems}, Oxford Univ. Press, Oxford (1997);
M. Sahimi, {\it Heterogeneous Materials II: Nonlinear Breakdown Properties and 
Atomistic Modelling}, Springer-Verlag Heidelberg, (2003);
H. J. Herrmann and S. Roux, {\it Statistical Models of Disordered Media}, North 
Holland, Amsterdam (1990).

\bibitem{zapperi97}

R. L. Smith, Proc. R. Soc. London A {\bf 372}, 539 (1980);
S. Zapperi, P. Ray, H. E. Stanley, A. Vespignani, Phys. Rev. Lett. {\bf 78}, 1408 (1997);
J. V. Andersen, D. Sornette and K. T. Leung, Phys. Rev. Lett {\bf 78}, 2140
(1997); S. D. Zhang and E-jiang Ding, Phys. Rev. {\bf 53}, 646 (1996); B. Q. Wu and
P. L. Leath,  Phys. Rev. B {\bf 59}, 4002 (1999).


\bibitem{silveria}
F. T. Peirce, J. Text. Inst. {\bf 17}, 355 (1926);
H. E. Daniels, Proc. R. Soc. London A {\bf 183} 404 (1945);
B. D. Coleman, J. Appl. Phys. {\bf 29}, 968 (1958);
R. L. Smith, Proc. R. Soc. London A {\bf 372}, 539 (1980);
R. da Silveria, Am. J. Phys. {\bf 67}, 1177 (1999);


\bibitem{hemmer}
P. C. Hemmer and Alex Hansen, J. Appl. Mech. {\bf 59}, 909 (1992);
A. Hansen and P. C. Hemmer, Trends in Stat. Phys. {\bf 1}, 213 (1994);
A. Hansen and P. C. Hemmer, Phys. Lett. A {\bf 184}, 394 (1994).

\bibitem{aval}
M. Kloster, A. Hansen and P. C. Hemmer, Phys. Rev. E {\bf 56}, 2615 (1997).


\bibitem{moreno93}
J. B. Gomez, D. Iniguez and A. F. Pacheco, Phys. Rev. Lett. {\bf 71}, 380 (1993). 

\bibitem{andersen}
J. V. Andersen, D. Sornette, and K. -T. Leung, Phys. Rev. Lett. {\bf 78}, 2140 (1997);
D. Sornette and J. V. Andersen, Eur. Phys. J. B {\bf 1}, 353 (1998).

\bibitem{moreno00}
Y. Moreno, J. B. Gomez and A. F. Pacheco, Phys. Rev. Lett. {\bf 85}, 2865 (2000).

\bibitem{reviewchak}
S. Pradhan and B. K. Chakrabarti, Int. J. Mod. Phys. B {\bf 17}, 5565 (2003);
P. C. Hemmer, A. Hansen and S. Pradhan, e-print cond-mat/0602371; in {\it 
Modelling Critical and Catastrophic Phenomena in Geoscience}, edited by
P. Bhattacharya and B. K. Chakrabarti (Springer, Berlin, 2006). p. 27.

\bibitem{dynamic}
S. Pradhan, P. Bhattacharyya and B.K. Chakrabarti, Phys. Rev. E {\bf 66}, 016116
(2002);
P. Bhattacharyya, S. Pradhan and B.K. Chakrabarti, Phys. Rev. E {\bf 67},
046122 (2003).

\bibitem{prl}
S. Pradhan, A. Hansen and P. C. Hemmer, Phys. Rev. Lett. {\bf 95}, 125501 
(2005); S. Pradhan, A. Hansen and P. C. Hemmer, Phys. Rev. E {\bf 74} 016122
(2006);
S. Pradhan and A. Hansen, Phys. Rev. E. {\bf 72}, 026111 (2005).

\bibitem{kim05} D. -H. Kim, B. J. Kim and H. Jeong, Phys. Rev. Lett.{\bf 94}, 025501 (2005);  U. Divakaran and A. Dutta, Int. J. Mod. Phys C, {\bf 18} 919 (2007)

\bibitem{kun06} 
F. Raischel, F. Kun and H. J. Herrmann, Phys. Rev. E {\bf 74}, 035104(R) (2006);
 F. Kun and S. Nagy, Phys. Rev. E {\bf 77}, 016608 (2008).
\bibitem{divakaran071} Uma Divakaran and Amit Dutta, Phys. Rev. E {\bf 75} 011109 (2007).

\bibitem{divakaran07} Uma Divakaran and Amit Dutta, Phys. Rev. E, {\bf 75} 
011117 (2007).

\bibitem{pradhan07}
S. Pradhan and Per C. Hemmer, Phys. Rev. E {\bf 77}, 031138 (2008);
S. Pradhan and Per C. Hemmer, Phys. Rev. E {\bf 75}, 056112 (2007);
Per C. Hemmer and S. Pradhan, Phys. Rev. E {\bf 75}, 046101 (2007).

\bibitem{hidalgo08} R. C. Hidalgo, K. Kovacs, I. Pagonabarraga and F. Kun,
 Eur. Phys. Lett., {\bf 81} 54005 (2008). 






\end{thebibliography}
\end{document}